\documentclass[reprint,amsmath,amssymb]{revtex4-2}

\usepackage{graphicx,amsmath,amsfonts}% Include figure files
\usepackage{amssymb}

\newcommand{\Af}{\hat{\mathbf{A}}_f}
\newcommand{\Ef}{\hat{\mathbf{E}}_f}
\newcommand{\Pf}{\hat{\mathbf{P}}_f}

\newcommand{\eps}{\boldsymbol{\varepsilon}}

\newcommand{\rs}{(\mathbf{r};s)}
\newcommand{\Phat}{\hat{\boldsymbol{\mathcal{P}}}}

\newcommand{\Dhat}{\hat{\boldsymbol{\mathcal{D}}}}

\newcommand{\tp}{\left( t \right) }

\newcommand{\hPib}{\hat{\boldsymbol{\Pi}}}

\newcommand{\Pib}{\boldsymbol{\Pi}}

\newcommand{\hQb}{\hat{\mathbf{Q}}_\nu}

\newcommand{\Ab}{\mathbf{A}}

\newcommand{\hA}{\hat{\mathbf{A}}}

\newcommand{\hYb}{\hat{\mathbf{Y}}}

\newcommand{\I}{\tensor{\mathbf{I}}}

\newcommand{\A}{\mathbf{A}}

\newcommand{\rb}{\mathbf{r}}
\newcommand{\rp}{\left( \mathbf{r} \right)}

\newcommand{\rpt}{\left( \mathbf{r};t \right)}
\newcommand{\dV}{\text{d}^3{\bf r}}
\newcommand{\dk}{\text{d}^3{\bf k}}
\newcommand{\dS}{\text{d}^2{\bf r}}

\newcommand{\rt}{(\mathbf{r};t)}

\makeatletter
\newsavebox{\@brx}
\newcommand{\llangle}[1][]{\savebox{\@brx}{\(\m@th{#1\langle}\)}%
  \mathopen{\copy\@brx\kern-0.5\wd\@brx\usebox{\@brx}}}
\newcommand{\rrangle}[1][]{\savebox{\@brx}{\(\m@th{#1\rangle}\)}%
  \mathclose{\copy\@brx\kern-0.5\wd\@brx\usebox{\@brx}}}

\makeatother

\begin{document}
%\preprint{APS/123-QED}

\title{Integral Formulation of Macroscopic Quantum Electrodynamics \\
in Dispersive Dielectric Objects}

% Force line breaks with \\
\author{Carlo Forestiere}
\affiliation{Department of Electrical Engineering and Information Technology, Universit\`{a} degli Studi di Napoli Federico II, via Claudio 21,
 Napoli, 80125, Italy}
\author{Giovanni Miano}
\affiliation{ Department of Electrical Engineering and Information Technology, Universit\`{a} degli Studi di Napoli Federico II, via Claudio 21, Napoli, 80125, Italy}
\begin{abstract}

We propose an integral formulation of macroscopic quantum electrodynamics in the Heisenberg picture for linear dispersive dielectric objects of finite size, utilizing the Hopfield-type approach. By expressing the electromagnetic field operators as a function of the polarization density field operator via the retarded Green function for the vacuum, we obtain an integral equation that governs the evolution of the polarization density field operator. This formulation offers significant advantages, as it allows for the direct application of well-established computational techniques from classical electrodynamics to perform quantum electrodynamics computations in open, dispersive, and absorbing environments.

 %We present an integral formulation of macroscopic quantum electrodynamics in the Heisenberg picture for linear dispersive dielectric objects of finite sizes using the Hopfield-type approach. We express the electromagnetic field operators as function of the polarization density field operator through the retarded Green's function for the vacuum, and we derive the integral equation that governs the evolution of the polarization density field operator. The proposed formulation is particularly attractive because it enables the direct application of consolidated repertory in computational classical electrodynamics to carry out quantum electrodynamics computation in open, dispersive, and absorbing environments.%
\end{abstract}
\maketitle

%\tableofcontents

\section{Introduction and main results}
 Macroscopic quantum electrodynamics in open, dispersive and
absorbing environments (e.g., \cite{glauber_quantum_1991, huttner_quantization_1992}) are of fundamental importance
in nanoplasmonics, nanophotonics (e.g., \cite{tame_quantum_2013, flamini_photonic_2018, pres_detection_2023}), and superconducting circuits (e.g., \cite{blais_circuit_2021}). Dispersion and absorption can be described by two types of approaches: the Hopfield-type approach \cite{{hopfield_theory_1958}} in which the matter is described as a harmonic oscillating bosonic field
linearly coupled to the electromagnetic field (e.g., \cite{huttner_quantization_1992}, \cite{suttorp_field_2004, bhat_hamiltonian_2006, philbin_canonical_2010}); the Langevin-noise approach based on a phenomenological noise current operator \cite{matloob_electromagnetic_1995, gruner_correlation_1995,vogel_quantum_2006}, which is applied in many contexts (e.g., \cite{scheel_macroscopic_2008, franke_quantization_2019}). These approaches are equivalent if in the Langevin-noise approach the quantized photonic degrees of freedom associated with the fluctuating radiation field are added to the degrees of freedom of the material oscillators (e.g., \cite{drezet_equivalence_2017,dorier_critical_2020}). 

The interaction between electromagnetic fields and dielectric objects is typically studied by diagonalizing the Hamiltonian operator. Diagonalization requires either the Green function in the presence of the dielectric object (e.g. \cite{dung_three-dimensional_1998}) or the eigenmodes of the dielectric object (e.g. \cite{glauber_quantum_1991,bhat_hamiltonian_2006}). 
A numerical mode decomposition for diagonalizing
the Hamiltonian has been proposed
\cite{na_diagonalization_2021}. Nevertheless, diagonalization of the Hamiltonian is not necessary. Recently, it has been proposed to expand the polarization density field operator in terms of the quasistatic modes of the dielectric object \cite{forestiere_operative_2022,forestiere_time-domain_2021}, where few quasistatic modes are required for small-sized dielectric objects.

In the Heisenberg picture, the macroscopic quantum electrodynamics of dielectric objects can be effectively studied by first evaluating the time evolution of the polarization density field operator and then the electric and magnetic field operators it generates. The separation between matter and electromagnetic field allows us to treat on an
equal footing both electromagnetic field and matter fluctuations.

\begin{figure}
    \centering
    \includegraphics[width=0.75\columnwidth]{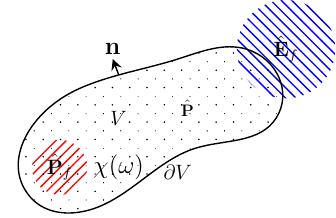}
    \caption{Region $V$ occupied by the dielectric object with susceptibility $\chi \left(\omega\right)$ and boundary $\partial V$. The dielectric is described by the Hopfield-type model. The temporal evolution of the polarization density field operator is driven by the free electric field operator $\hat{\mathbf{E}}_f\rpt$ and the free polarization field operator $\hat{\mathbf{P}}_f\rpt$. They take into account the influence of the initial condition of the matter and radiation field operators, through which the initial quantum state of the system comes into play. They are linear combinations of the annihilation and creation operators, in the
Schr\"odinger picture, for the radiation and matter fields.}
    \label{fig:Fig1}
\end{figure}

In this paper, we introduce an integral equation for the polarization density field operator in the Heisenberg picture that is based on the retarded Green function for the vacuum. We consider a linear, isotropic, dispersive dielectric object of finite size (Fig. \ref{fig:Fig1}): $\chi(\omega)$ denotes the macroscopic susceptibility of the dielectric
in the frequency domain, $V$ is the region of space occupied by the object, $\partial V$ its boundary, $\mathbf{n}$ the unit vector normal to $\partial V$ that points outward and $V_\infty$ the overall unbounded space. We use a Hopfield-type approach to describe the dielectric \cite{{hopfield_theory_1958}}.
We find that the polarization density field operator $\hat{\mathbf{P}}\rt$ is a solution (for $t>0$) of the linear volume integral equation
\begin{equation}
\label{eq:Pol}
h_\eta (t)*\hat{\mathbf{P}}\rt -{\varepsilon_0 \,}\mathcal{L}\{\hat{\mathbf{P}}\}\rt = \hat{\mathbf{D}}_f\rt \quad \text{in}\, V.
\end{equation}
The symbol $*$ denotes the time convolution product over the interval $[0,t]$, $h_\eta (t)$ is the inverse Fourier transform of $\eta(\omega)=1/\chi(\omega)$,
\begin{widetext}
\begin{equation}
    \label{eq:OperatorL}
    \mathcal{L}\{\hat{\mathbf{P}}\}\rt =-\frac{\partial}{\partial t}\left[\frac{\mu_0}{4 \pi}\int_V\frac{{\dot{\hat{{\mathbf{P}}}}}(\mathbf{r}';t')}{\left| {\bf r} - {\bf r}' \right|}\dV'\right] - \nabla \left[ \frac{1}{4 \pi \varepsilon_0} \int_{\partial V} \,\frac{{{\hat{\mathbf{P}}}}(\mathbf{r}';t')\cdot\,\mathbf{n}(\mathbf{r'})}{\left| {\bf r} - {\bf r}' \right|} \dS'\right],
\end{equation}
\end{widetext}
$t' = t-|\mathbf{r}-\mathbf{r}'|/c_0$ is the retarded time between the points $\mathbf{r}$ and $\mathbf{r}'$, the dot denotes the first derivative with respect to the time, ${\dot{\hat{\mathbf{P}}}=\hat{\mathbf{J}}\rt}$ is the polarization current density operator, $\mu_0$ is the vacuum permeability, $\varepsilon_0$ is the vacuum permittivity, and $c_0=\sqrt{1/\varepsilon_0 \mu_0}$.
The field operator $\hat{\mathbf{D}}_f\rt$ is known and takes into account the contribution of the initial conditions of the matter and radiation field operators, through which the initial quantum state of the system comes into play. It is given by
\begin{equation}
\label{eq:Ftot}
  \hat{\mathbf{D}}_f\rt =h_\eta (t)*\Pf\rt+\varepsilon_0\hat{\mathbf{E}}_f\rpt,
\end{equation}
where $\hat{\mathbf{E}}_f\rpt$ and $\Pf\rt$ are given, respectively, by Eqs. \ref{eq:Efree_t} and \ref{eq:Pfree1t}. 
The equation \ref{eq:Pol} must be solved with the initial condition $\mathbf{\hat{{P}}}(\mathbf{r};t=0)=\hat{\mathbf{P}}^{(S)}(\mathbf{r})$ where the upper script $^{(S)}$ denotes the operator in the Schr\"odinger picture.
The operators $\Pf\rt$ and $\hat{\mathbf{E}}_f\rpt$ are linear combinations of the annihilation and creation operators, in the Schr\"odinger picture, for the bosonic matter fields and the radiation field. 

The electric field operator $\hat{\mathbf{E}}\rt$ is given in $V_\infty$ and for $t\ge 0$, by
\begin{equation}
\label{eq:Eoper}
\hat{\mathbf{E}}\rt = \mathcal{L}\{\hat{\mathbf{P}}\}\rt + \hat{\mathbf{E}}_f\rt.
\end{equation}
The magnetic field operator $\hat{\mathbf{B}}\rt$ is given by
\begin{equation}
\label{eq:Boper}
    \hat{\mathbf{B}}\rt = \nabla \times \left[\frac{\mu_0}{4\pi} \int_V\frac{{\dot{\hat{\mathbf{P}}}}(\mathbf{r}';t')}{\left| {\bf r} - {\bf r}' \right|}\dV'\right] + \hat{\mathbf{B}}_f\rt
\end{equation}
where $\hat{\mathbf{B}}_f=\nabla \times \Af$ takes into account the contribution of the initial conditions of the radiation field operators; the expression of $\Af$ is given by \ref{eq:Afree}. In the expressions \ref{eq:OperatorL} and \ref{eq:Boper} the operators $\hat{\mathbf{P}}$ and $\dot{\hat{\mathbf{P}}}$ are extended to $t<0$ in such a way as to be null. 

Equations \ref{eq:Pol}-\ref{eq:Boper} are the main results of this paper. This formulation is extremely attractive because the consolidated repertoire of computational classical electrodynamics can be used to solve the equation \ref{eq:Pol} and to evaluate the expressions \ref{eq:Eoper} and \ref{eq:Boper}. Specifically, the vector field operator $\hat{\mathbf{D}}_f$ is a linear combination of the annihilation operators $\hat{{c}}_{m,\nu}$ and the creation operators $\hat{{c}}_{m,\nu}^\dagger$ for the bosonic matter fields, and of the annihilation operators $\hat{{a}}_{\mu}$ and the creation operators $\hat{{a}}_{\mu}^\dagger$ for the radiation field (in the Schr\"odinger picture). Therefore, Eq. \ref{eq:Pol} can be solved by expanding the polarization density field operator in terms of the operators $\hat{{c}}_{m,\nu}, \,\hat{{c}}_{m,\nu}^\dagger, \, \hat{{a}}_{\mu}$ and $\hat{{a}}_{\mu}^\dagger$. The expansion coefficients are c-vector fields, that is, real or complex vector fields (which are not operators). They are solutions of integral equations of the type (see Eq. \ref{eq:IntegralEquation}) encountered in classical electromagnetic scattering (e.g., \cite{van_bladel_electromagnetic_2007}), which can be solved numerically by finite elements, either by resorting to marching-on-time techniques or by frequency domain analysis and Fourier transform. In the same way, the expressions \ref{eq:Eoper} and \ref{eq:Boper} can be evaluated.

In Section II, we summarize the Hamiltonian formulation of the interaction between a dispersive dielectric object of finite size and the electromagnetic field in the framework of macroscopic quantum electrodynamics. In Section III, we first determine the Heisenberg equations for the fundamental observables of the system, then we solve them and derive Eqs. \ref{eq:Pol}-\ref{eq:Boper}. In Section IV, we show that Eq. \ref{eq:Pol} can be recast into a set of volume integral equations whose unknowns are c-vector fields, and expressions \ref{eq:Eoper} and \ref{eq:Boper} can be recast into a set of integrals of c-vector fields. In Section V, we face the problem of evaluating the statistical functions of the system. In Section VI, we discuss the choice of the Coulomb gauge. Finally, we present conclusions on the proposed approach in Section VII.

\section{Hamiltonian formulation}

We consider a linear, isotropic, and dispersive dielectric object of finite size that interacts with the electromagnetic field. To simplify the presentation, we assume that the dielectric object is homogeneous.

\subsection{Classical Hamiltonian}

We use the Hopfield-type approach \cite{{hopfield_theory_1958}} to model the electromagnetic
field that interacts with the dispersive dielectric. In the framework of classical electrodynamics, the dielectric is represented by a continuum of harmonic oscillating fields $\boldsymbol{Y}_\nu\rt$ defined in $V$, where $\nu$ is the natural frequency of the oscillating field with $0\le\nu<\infty$ (\cite{bhat_hamiltonian_2006}, \cite{philbin_canonical_2010}, \cite{forestiere_operative_2022}). The polarization density field $\boldsymbol{P}\rpt$ is expressed as
\begin{equation}
   \label{eq:Prel}
    \mathbf{P}\rpt = \left\{
    \begin{array}{cl}
        \int_0^\infty \alpha_\nu \boldsymbol{Y}_\nu\rpt  d\nu & \text{in} \, V,  \\
          0 & \text{in} V_\infty \backslash V,
    \end{array}
    \right.
\end{equation}
where $\alpha_\nu = \sqrt{2{\sigma}(\nu)/{\pi}}$ is the coupling parameter, $\sigma(\omega) = -\varepsilon_0 \omega{\chi}_i(\omega)$, and $\chi_i(\omega)$ is the imaginary part of the dielectric susceptibility. 
Due to the homogeneity hypothesis, the coupling parameter is independent of the space variable, the vector field $\boldsymbol{P}$ is solenoidal in $V$ and has a non-vanishing normal component to $\partial V$, which is the surface polarization charge density induced on the dielectric surface $P_{n}\rt=\boldsymbol{P} \cdot \mathbf{n}$.

The electromagnetic field is represented through the electromagnetic potentials by using the Coulomb gauge. As a consequence, the electric field is decomposed into longitudinal and transverse components (Appendix A). The longitudinal component is not a true dynamical variable of the problem since it depends on the polarization charge through the Coulomb law. Therefore, the degrees of freedom of the entire system are the continuous set of matter fields $\{\boldsymbol{Y}_\nu\rt\}$ and the vector potential $\boldsymbol{A}\rt$. 
The canonically conjugate variables of the continuum of matter fields are $\boldsymbol{Y}_\nu\rt$ and its conjugate momentum $\boldsymbol{Q}_\nu \rpt=\dot{\boldsymbol{Y}}_\nu +\alpha_\nu\boldsymbol{A}$. The canonically conjugate variables of the radiation field are the vector potential $\boldsymbol{A}\rt$ and its conjugate momentum $\boldsymbol{\Pi}\rpt=\varepsilon_0 \dot{\boldsymbol{A}}$ (e.g., \cite{suttorp_field_2004}, \cite{forestiere_operative_2022}). 

The Hamiltonian of the entire system has three contributions, ${H} = {H}_{mat}+{H}_{Coul}+{H}_{rad}$. The matter energy term ${H}_{mat}(\boldsymbol{Q}_\nu,\boldsymbol{Y}_\nu,\boldsymbol{A})$, the Coulomb interaction energy term ${H}_{Coul}(\boldsymbol{P})$, and the radiation energy term ${H}_{rad}(\boldsymbol{\Pi},\boldsymbol{A})$ have the following expressions (e.g., \cite{suttorp_field_2004}, \cite{forestiere_operative_2022}):
\begin{subequations}
\begin{align}
\label{eq:hmat}
   {H}_{mat} &= \int_V \dV \int_0^\infty d\nu \left [\frac{1}{2} \left( \boldsymbol{Q}_\nu - \alpha_\nu {\boldsymbol{A}} \right)^2+\frac{\nu ^2} {2}\boldsymbol{Y}_\nu^2\right],\quad \\
\label{eq:hcou}
   {H}_{Coul} &= \int_{\partial V}\dS \int_{\partial V}\dS'\, \frac{P_{n} \rt P_{n} (\mathbf{r'};t)}{8\pi\varepsilon_0|\mathbf{r}-\mathbf{r}'|},\\
\label{eq:hrad}
   {H}_{rad} &= \int_{V_\infty} \dV \left[ \frac{1}{2\varepsilon_0} \Pib^{2} + \frac{1}{2\mu_0} \left(\nabla\times {{\boldsymbol{A}}} \right)^2 \right].
\end{align}
\end{subequations}

\subsection{Quantization}
 
 In the Heisenberg picture, the vector field operators $\hQb\rt$ and $\hYb_\nu\rt$ correspond to the canonically conjugate variables $\boldsymbol{Q}_\nu\rt$ and $\boldsymbol{Y}_\nu\rt$, and the vector field operators $\hPib\rt$ and $\hA\rt$ correspond to the canonically conjugate variables $\Pib\rt$ and $\boldsymbol{A}\rt$. They obey the equal-time commutation relations (e.g., \cite{cohen-tannoudji_photons_1997}, \cite{suttorp_field_2004}, \cite{philbin_canonical_2010})
\begin{equation}
\left[ \hQb\rt, \hYb_{\nu'} (\mathbf{r'};t) \right]  =  -i \hbar \I \delta \left( \nu - \nu' \right) \boldsymbol\delta \left( \rb - \rb' \right)  \text{\;in } V,
\end{equation}
\begin{equation}
\left[ \hPib \rt, \hA (\mathbf{r'};t) \right] = -i \hbar \boldsymbol\delta^\perp \left( \rb - \rb' \right)   \text{\;in } V_\infty,
\end{equation}
where $\boldsymbol\delta^\perp$ is the transverse dyadic delta function, Appendix A. All remaining equal-time commutators vanish.

The vector field operator $\hat{\mathbf{P}}\rt$ corresponds to the polarization density field $\boldsymbol{P}\rt$,
\begin{equation}
\label{eq:Polq0}
\hat{\mathbf{P}}\rt = \int_0^\infty \alpha_\nu
\hat{\mathbf{Y}}_\nu\rt d\nu \quad \text{in}\, V
\end{equation}
and it is null outside $V$. Its conjugate momentum is given by
\begin{align}
\label{eq:Polq}
\hat{\mathbf{Z}}\rt &=\frac{1}{\beta} \int_0^\infty \alpha_\nu
{\hat{\mathbf{Q}}}_\nu\rt d\nu \quad \text{in}\, V
\end{align}
where $\beta=\int_0^\infty \alpha_\nu^2\, d\nu$; it is null outside $V$.
They obey the equal time commutation relation
\begin{equation}
\label{eq:commut_1}
\left[ \hat{\mathbf{Z}}\rt, \hat{\mathbf{P}}(\mathbf{r'};t) \right]  =  -i\hbar \I \delta \left( \rb - \rb' \right) \text{in}\, V.
\end{equation}
The other equal time commutators with $\hat{\bf A}$ and $\hPib$ vanish. The polarization current density operator $\hat{\mathbf{J}}\rt$
is given by $\hat{\mathbf{J}}=\beta\left(\hat{\mathbf{Z}} - \hat{\bf A}\right)$ in $V$ and is null outside $V$.

The electric field operator $\hat{\mathbf{E}}\rt$ is given by 
\begin{align}
\label{eq:Eop}
\hat{\mathbf{E}}=-\frac{1}{\varepsilon_0}{\hat{\mathbf{\Pi}}} +{\mathbf{E}}^\parallel\{\hat{\mathbf{P}}\}
\end{align}
where
\begin{equation}
\label{eq:Elong}
    \mathbf{E}^\parallel\{\hat{\mathbf{P}}\} =-\frac{1}{\varepsilon_0} \nabla\int_{\partial V} \,\frac{ \hat{{\mathbf{P}}} \left( \rb ';t\right)\cdot\,\mathbf{n}(\mathbf{r'})}{4\pi\left| {\bf r} - {\bf r}' \right|}\dS'.
\end{equation}
The term $-{\hat{\mathbf{\Pi}}}/{\varepsilon_0}$ is the transverse component of $\hat{\mathbf{E}}$ and ${\mathbf{E}}^\parallel\{\hat{\mathbf{P}}\}$ is the longitudinal component. The vector field operator $\mathbf{E}^\parallel\{\hat{\mathbf{P}}\}$ is solenoidal in $V$ and $V_\infty \backslash V$, but its normal component to $\partial V$ is discontinuous. We note that $\varepsilon_0 \mathbf{E}^\parallel\{\hat{\mathbf{P}}\}={\hat{\mathbf{P}}}^\parallel$ in $V_\infty$. 

The magnetic field operator $\hat{\mathbf{B}}\rt$ is given by 
\begin{equation}
\label{eq:B_VP}
    \hat{\mathbf{B}}=\nabla\times \hat{\mathbf{A}}.
\end{equation}
\\
\section{Heisenberg equations}

The Heisenberg equations for the canonical variables of the problem follow by evaluating the corresponding commutators with the quantized Hamiltonian. 

The conjugate canonical operators $\hat{\mathbf{Y}}_\nu$ and $\hat{\mathbf{Q}}_\nu$, for $0\le\nu<\infty$, are governed for $0<t$ by the system of equations defined in $V$
\begin{subequations}
\label{eq:YQ}
\begin{eqnarray}
    \label{eq:Yc}
    \dot{\hat{\mathbf{Y}}}_\nu&=& \hat{\mathbf{Q}}_\nu - \alpha_\nu{\hat{\bf A}}, \quad \\
    \label{eq:Qc}
    \dot{\hat{\mathbf{Q}}}_\nu &=&-\nu^2\hat{\mathbf{Y}}_\nu+\alpha_\nu \mathbf{E}^\parallel\{\hat{\mathbf{P}}\}.
\end{eqnarray}
\end{subequations}
 The canonical conjugate operators $\hat{\mathbf{A}}$ and $\hat{\mathbf{\Pi}}$ are governed for $0<t$ by the system of equations defined in $V_\infty$
\begin{subequations}
\label{eq:emperp}
\begin{eqnarray}
    \label{eq:Aperpc} 
    \dot{\hat{\A}} &=&\frac{1}{\varepsilon_0} \hat{\Pib} \;, \\
    \label{eq:Pperpc}
    \dot{\hat{\Pib}} &=& \frac{1}{\mu_0} \nabla^2 \hat{\A} + \dot{\hat{\mathbf{P}}}^\perp,
\end{eqnarray}
\end{subequations}
where ${\hat{\mathbf{P}}}^\perp = {\hat{\mathbf{P}}}-\varepsilon_0 \mathbf{E}^\parallel\{\hat{\mathbf{P}}\}$ is the transverse component of the polarization density field operator.
Eliminating the operators $\hat{\mathbf{Q}}_\nu$ and $\hat{\mathbf{\Pi}}$, the system of equations \ref{eq:YQ} and \ref{eq:emperp} reduce to
\begin{subequations}
\begin{align}
\label{eq:Adotdot}
    \ddot{\hat{\A}} - c_0^2 \nabla^2 \Hat{\A} &= \frac{1}{\varepsilon_0} \dot{\hat{\mathbf{P}}}^\perp \text{\;in } V_\infty,\\
\label{eq:Ydotdot}
    \ddot{\hat{\mathbf{Y}}}_\nu + \nu^2 {\hat{\mathbf{Y}}}_\nu &=\alpha_\nu[ -\dot{\hat{\A}}+\mathbf{E}^\parallel\{\hat{\mathbf{P}}\}] \text{\;in } V.
\end{align}
\end{subequations}
This system of equations has to be solved with the initial conditions: 
\begin{subequations}
\begin{align}
\label{eq:inita}
    \hat{\mathbf{A}}(\mathbf{r};t=0)&=\hat{\mathbf{A}}^{(S)}(\mathbf{r}),\\
\label{eq:initdotA}
    \dot{\hat{\mathbf{A}}}(\mathbf{r};t=0)&=\frac{1}{\varepsilon_0}\hat{\mathbf{\Pi}}^{(S)}(\mathbf{r}),\\
\label{eq:inity1}
\hat{\mathbf{Y}}_\nu(\mathbf{r};t=0)&=\hat{\mathbf{Y}}_\nu^{(S)}(\mathbf{r}),\\
\label{eq:inity2}
\dot{\hat{\mathbf{Y}}}_\nu(\mathbf{r};t=0)&=\dot{\hat{\mathbf{Y}}}_\nu^{(S)}(\mathbf{r})=\hat{\mathbf{Q}}_\nu^{(S)}(\mathbf{r})-\alpha_\nu\hat{\mathbf{A}}^{(S)}(\mathbf{r}).
\end{align}
\end{subequations}
We recall that the upper
script (S) denotes the operator in the
Schr\"odinger picture. 

The procedure for obtaining equations \ref{eq:Pol}-\ref{eq:Boper} is straightforward. The solution of Eq. \ref{eq:Adotdot} with the initial conditions \ref{eq:inita} and \ref{eq:initdotA} gives $\hat{\mathbf{A}}$ as a function of $\hat{\mathbf{P}}$ through the transverse dyadic Green function for vacuum. Then, using \ref{eq:Eop} and \ref{eq:Aperpc} we obtain the expression \ref{eq:Eoper} for the electric field operator where $\mathcal{L}\{\hat{\mathbf{P}}\}$ is given by \ref{eq:OperatorL}; using \ref{eq:B_VP} we obtain the expression \ref{eq:Boper} for the magnetic field operator. The solution of Eq. \ref{eq:Ydotdot} with the initial conditions \ref{eq:inity1} and \ref{eq:inity2} gives the constitutive relation of the dielectric. Substituting into the constitutive relation the expression of the electric field operator given by \ref{eq:Eoper} we obtain Eq. \ref{eq:Pol}. In the following, we provide the details with the support of Appendices B, C, and D.

\subsection{Vector potential operator}

Solving equation \ref{eq:Adotdot} with the initial conditions \ref{eq:inita} and \ref{eq:initdotA} we obtain the expression of the vector potential operator $\hat{\mathbf{A}}\rt$ for $t>0$ (for details, see Appendix B),
\begin{equation}
\label{eq:Atime}
    \hat{\mathbf{A}}\rt =\mu_0 \int_V\dV' {\overleftrightarrow{g}}^\perp \left(\mathbf{r} - \mathbf{r}';t\right)* \dot{\hat{\mathbf{P}}}(\mathbf{r}';t)  + \Af\rt
\end{equation}
where the symbol $*$ denotes the time convolution product over the interval $[0,t]$, and $\overleftrightarrow{g}^\perp \left(\rb;t \right)$ is the transverse dyadic Green function whose expression is given in Appendix B by \ref{eq:GreenTravTime0}. The \textit{free vector potential operator} $\Af\rt$ takes into account the contribution of the initial conditions of the radiation field operators, $\mathbf{\hat{{A}}}_f(\mathbf{r};t=0)=\hat{\mathbf{A}}^{(S)}(\mathbf{r})$. 

The dyad $\overleftrightarrow{g}^\perp$ is the difference between the Green function for the vector potential in the temporal gauge (that is, the retarded Green function for the electric field) and its longitudinal component (that is, the Green function for the electrostatic field, except for the factor $c_0^2u(t)t$ where $u(t)$ is the Heaviside function).

To represent $\mathbf{\hat{{A}}}_f$ we use the transverse plane wave modes of the free space
\begin{equation}
\label{eq:planewave}
    \mathbf{w}_\mu \rp = \frac{1}{\left(2\pi\right)^{3/2}} \eps_{s,\mathbf{k}} e^{i \mathbf{k} \cdot \rb},
\end{equation}
where ${\bf k} \in \mathbb{R}^3$ is the propagation vector,  $\left\{ \eps_{s,\mathbf{k}}\right\}$ are the unit vectors of polarization with $\eps_{s,\mathbf{k}} = \eps_{s,-\mathbf{k}}$ and $s=1,2$; $\mu = \left(\mathbf{k},s \right)$ is a multi-index corresponding to the pair of parameters $\mathbf{k}$ and $s$, $\mathcal{M}$ represents the set of all possible $\mu$, and $\sum_\mu \left( \cdot \right) = \displaystyle\sum_{\mu \in \mathcal{M}} \int_{\mathbb{R}^3} \dk\,\left( \cdot \right) $.
The two polarization vectors are orthogonal among them, $\eps_{1,\mathbf{k}} \cdot \eps_{2, \mathbf{k}} = 0$, and both are transverse to the propagation vector, $\eps_{1,\mathbf{k}} \cdot \mathbf{k} = \eps_{2, \mathbf{k}} \cdot \mathbf{k}=0$.
The functions $\left\{ \mathbf{w}_\mu \right\}$ are orthonormal, i.e. $\langle \mathbf{w}_{\mu'}, \mathbf{w}_\mu  \rangle = \delta_{s',s} \delta \left( \mathbf{k} - \mathbf{k}' \right)$. The free vector potential operator is given by
\begin{equation}
\label{eq:Afree}
\Af\rpt =\sum_\mu\frac{1}{i\omega_\mu}\boldsymbol{e}_\mu\rt \hat{a}_\mu+h.c.,
\end{equation}
where
\begin{equation}
    \label{eq:e_mu}
    {\boldsymbol{e}}_\mu\rpt =\mathcal{E}_\mu \mathbf{w}_\mu\rp e^{-i\omega_\mu t},
\end{equation}
$\omega_\mu=c_0k$, $\mathcal{E}_\mu=\{\hbar \omega_\mu/[2\varepsilon_0 (2\pi)^3]\}^{1/2}$ and h.c. is the abbreviation for the ``hermitian conjugate''. 
The operators $\hat{a}_\mu$ and $\hat{a}^\dagger_\mu$ are the annihilation and creation operators in the Schr\"odinger picture for the transverse electromagnetic field mode with index $\mu$. They obey the commutation relations $[\hat{a}_\mu,\hat{a}_\mu'^\dagger]=\delta(\mu-\mu')$, while all other commutators vanish. The expression of $\hat{\mathbf{A}}^{(S)}$ in terms of the annihilation and creation operators is given by 
\begin{equation}
\label{eq:As}
\hat{\mathbf{A}}^{(S)}(\mathbf{r})=\sum_\mu\frac{1}{i\omega_\mu}\mathcal{E}_\mu \mathbf{w}_\mu\rp \hat{a}_\mu+h.c. \,.
\end{equation}
\\
\subsection{Dielectric constitutive relation}
 We obtain the constitutive relation of the dispersive dielectric by first solving equation \ref{eq:Ydotdot} with the initial conditions \ref{eq:inity1} and \ref{eq:inity2}, and then using the expression \ref{eq:Polq0} (for details, see Appendix C):
\begin{equation}
\label{eq:PolA1bis}
 \displaystyle \hat{\mathbf{E}}\left(\mathbf{r};t \right)  =\frac{1}{\varepsilon_0} h_\eta (t) * [ {\hat{\mathbf{P}}}\rt - 
 {\hat{\mathbf{P}}}_f\rt ]
 \text{\;in } V \text{\;and } 0<t
\end{equation}
where $h_\eta (t)$ is the inverse Fourier transform of $\eta(\omega)=1/\chi(\omega)$.

The \textit{free polarization density field operator} $\Pf\rt$ takes into account the contribution of the initial conditions of the matter field operators and of the vector potential operator, $\Pf(\mathbf{r};t=0)=\mathbf{\hat{{P}}}^{(S)}(\mathbf{r})$ where $\mathbf{\hat{{P}}}^{(S)}(\mathbf{r})$ is the polarization density field operator in the Schr\"odinger picture. To represent $\Pf$, we introduce an orthonormal discrete real basis $\{\mathbf{U}_m(\mathbf{r})\}$ defined in $V$. The operator $\mathbf{\hat{{P}}}_f$ is given by
\begin{equation}
\label{eq:Pfree1t}
\mathbf{\hat{{P}}}_f\rt = \mathbf{\hat{{P}}}_0\rt - \varepsilon_0 h_\chi (t)\hat{\mathbf{A}}^{(S)}(\mathbf{r}),
\end{equation}
where $h_\chi (t)$ is the inverse Fourier transform of $\chi(\omega)$,
\begin{equation}
\label{eq:Pfree10t}
\mathbf{\hat{{P}}}_0\rt = \sum_{m,\,\nu} \sqrt{\frac{\hbar \sigma(\nu)}{\nu \pi}} \, \mathbf{U}_m(\mathbf{r})\,e^{-i\nu t} \,\hat{{c}}_{m,\nu} + h.c.,
\end{equation}
and $\sum_{m, \, \nu} \left( \cdot \right)=\displaystyle\sum_{m} \int_0^\infty d\nu \,\left( \cdot \right) $. The operators $\hat{{c}}_{m,\nu}$ and $\hat{{c}}_{m,\nu}^\dagger$ are, respectively, the annihilation and creation operators in the Schr\"odinger picture for the matter field bosonic oscillators. They obey the commutation relations
$[\hat{c}_{m,\,\nu},\hat{c}_{m',\,\nu'}^\dagger]=\delta_{m,m'}\delta(\nu-\nu')$, and all the other commutators vanish.

Since $h_\chi(0)=0$ (Appendix C), we have $\mathbf{\hat{{P}}}^{(S)}(\mathbf{r})=\mathbf{\hat{{P}}}_0(\mathbf{r};t=0)$.

\subsection{Electric and magnetic field operators}
The electric field operator is given by \ref{eq:Eop} where the operator ${\hat{\mathbf{\Pi}}}$ is given by \ref{eq:emperp}, then $\hat{\mathbf{E}}=-{\dot{\hat{\mathbf{\A}}}} +{\mathbf{E}}^\parallel\{\hat{\mathbf{P}}\}$. In Appendix D, using the expression \ref{eq:Atime} for ${{\hat{\mathbf{\A}}}}$ and the expression \ref{eq:Elong} for ${\mathbf{E}}^\parallel\{\hat{\mathbf{P}}\}$ we show that
\begin{multline}
\label{eq:EApp1}
  \hat{\mathbf{E}}\rt = -\mu_0\frac{\partial}{\partial t}\int_V\frac{{\dot{\hat{\mathbf{P}}}}(\mathbf{r}';t')}{4\pi\left| {\bf r} - {\bf r}' \right|}u(t')\dV' - \\ \frac{1}{\varepsilon_0}\nabla\int_{\partial V} \,\frac{{\hat{\mathbf{P}}}(\mathbf{r}';t')\cdot\,\mathbf{n}(\mathbf{r'})}{4\pi\left| {\bf r} - {\bf r}' \right|} u(t')\dS'+\hat{\mathbf{E}}_{f}
\end{multline}
where $t' = t-|\mathbf{r}-\mathbf{r}'|/c_0$,
\begin{equation}
\label{eq:Efree_t}
\hat{\mathbf{E}}_f=\hat{\mathbf{E}}^\perp_f+\hat{\mathbf{E}}^\parallel_f,
\end{equation}
\begin{equation}
\label{eq:Efree_1}
\Ef^\perp \rpt =\sum_\mu\boldsymbol{e}_\mu\rpt\hat{a}_\mu+h.c.,
\end{equation}
\begin{equation}
\label{eq:Efree_3}
    \hat{\mathbf{E}}^\parallel_f\rt = \sum_{m,\,\nu} \sqrt{\frac{\hbar \sigma(\nu)}{\nu \pi}} \mathbf{N}_m\rt  \hat{{c}}_{m,\nu}+h.c.,
\end{equation}
\begin{multline}
\label{eq:Efree_4}
    \boldsymbol{N}_m(\mathbf{r};t)=
-\frac{1}{\varepsilon_0}\nabla \int_{\partial V}  \frac{{\mathbf{U}_m(\mathbf{r}')\cdot \mathbf{n}'\rp}}{4\pi |\mathbf{r}-\mathbf{r}'|}[u(t)-u(t')]\dS',
\end{multline}
and $u(t)$ is the Heaviside function. The expression \ref{eq:EApp1} can be rewritten as the expression \ref{eq:Eoper} where $\mathcal{L}\{\hat{\mathbf{P}}\}$ is given by \ref{eq:OperatorL}. The term $\hat{\mathbf{E}}^\perp_f$ takes into account the contribution due to the initial condition of the radiation field operators and $\hat{\mathbf{E}}^\parallel_f$ takes into account the contribution due to the initial condition of the polarization density field operators.
For the magnetic field operator, we obtain the expression \ref{eq:Boper} where $\hat{\mathbf{B}}_f=\nabla \times \Af$ (for the details, see Appendix D). We note that $\mathbf{\hat{{E}}}(\mathbf{r};t=0)=\mathbf{\hat{{E}}}_f(\mathbf{r};t=0)$ and $\mathbf{\hat{{B}}}(\mathbf{r};t=0)=\mathbf{\hat{{B}}}_f(\mathbf{r};t=0)$.

Combining equations \ref{eq:Eoper} and \ref{eq:PolA1bis} we obtain the integral equation \ref{eq:Pol} where $\hat{\mathbf{D}}_f\rt$ is given by \ref{eq:Ftot}, $\hat{\mathbf{P}}_f\rt$ is given by \ref{eq:Pfree1t} and $\hat{\mathbf{E}}_f\rt$ is given by \ref{eq:Efree_t}.
\\

\section{Expansion in terms of annihilation and creation operators}

The known term in Eq. \ref{eq:Pol}, $\hat{\mathbf{D}}_f \rt$, is a linear combination of the annihilation and creation operators for the matter fields and the transverse electromagnetic fields in the Schr\"odinger picture. Indeed, using the expression of $\hat{\mathbf{P}}_f \rt$ given by \ref{eq:Pfree1t}, \ref{eq:Pfree10t}, \ref{eq:As}, and the expression of $\hat{\mathbf{E}}_f \rt$ given by \ref{eq:Efree_t}-\ref{eq:Efree_4} we rewrite $\hat{\mathbf{D}}_f \rt$ as 
\begin{multline}
\hat{\mathbf{D}}_f\rt=\sum_\mu \boldsymbol{D}_\mu^{(rad)}\rt \hat{{{{a}}}}_\mu+ \sum_{m,\nu}\boldsymbol{D}_{m,\nu}^{(mat)}\rt \hat{{{{c}}}}_{m,\nu}+h.c.
\end{multline}
where $\boldsymbol{D}_\mu^{(rad)}$ and $\boldsymbol{D}_{m,\nu}^{(mat)}$ are c-vector fields given by
\begin{equation}
    \label{eq:Frad}
    \boldsymbol{D}_\mu^{(rad)}\rt= \varepsilon_0\mathcal{E}_\mu \mathbf{w}_\mu\rp\left( e^{-i\omega_\mu t}+\frac{i}{\omega_\mu}\right),
\end{equation}
and
\begin{multline}
\label{eq:Fmat}
\boldsymbol{D}_{m,\nu}^{(mat)}\rt=\sqrt{\frac{\hbar \sigma(\nu)}{\nu \pi}} \\ h_\eta (t)* \left\{\mathbf{U}_m(\mathbf{r})e^{-i\nu t}+ \boldsymbol{N}_m(\mathbf{r};t)\right\}.
\end{multline}
Since Eqs. \ref{eq:Pol}, \ref{eq:Eoper} and \ref{eq:Boper} are linear, the polarization density field operator, the electric field operator and the magnetic field operator are linear combinations of the annihilation and creation operators $\hat{{{{a}}}}_\mu$, $\hat{{{{a}}}}_\mu^\dagger$, $\hat{{{{c}}}}_{m,\nu}$ and $\hat{{{{c}}}}_{m,\nu}^\dagger$ in the Schr\"odinger picture.

\subsection{Volume integral equations for the c-vector fields}

We represent the polarization density field operator as
\begin{multline}
\label{eq:exp_P}
  \hat{\mathbf{{{P}}}}\rt=\sum_\mu \boldsymbol{p}_\mu^{(rad)}\rt \hat{{{{a}}}}_\mu + \sum_{m,\nu}\boldsymbol{p}_{m,\nu}^{(mat)}\rt \hat{{{{c}}}}_{m,\nu}+h.c.
\end{multline}
where the c-vector fields $\boldsymbol{p}_\mu^{(rad)}$ and $\boldsymbol{p}_{m,\nu}^{(mat)}$ are determined imposing Eq. \ref{eq:Pol}. Substituting \ref{eq:exp_P} into Eq. \ref{eq:Pol} we obtain that the c-vector field $\boldsymbol{{p}}_\beta ^{(\alpha)} \rt$ is solution for $0<t$ and in $V$ of the classical volume integral equation
\begin{equation}
\label{eq:IntegralEquation}
  h_\eta (t)*{\boldsymbol{p}_\beta^{(\alpha)}\rt - \varepsilon_0 \mathcal{L} \{\boldsymbol{p}}_\beta^{(\alpha)}\}\rt ={\boldsymbol{D}}_\beta^{(\alpha)}\rt
\end{equation}
where the expression of the linear integral operator $\mathcal{L}\{\mathbf{p}_\beta^{(\alpha)}\}$ is the same  as that given in \ref{eq:OperatorL}: in this equation the operator $\mathcal{L}\{\cdot\}$ acts on a c-vector field instead of a vector field operator. This equation has to be solved with the initial condition $\boldsymbol{p}_\mu^{(rad)}(\mathbf{r};0)=0$ when $\alpha = rad$, $\beta =\mu$, and with the initial condition $\boldsymbol{p}^{(mat)}_{m,\nu}(\mathbf{r};0)=\mathbf{U}_m(\mathbf{r})\sqrt{\frac{\hbar \sigma(\nu)}{\nu \pi}}$ when $\alpha = mat$, $\beta =m,\nu$.

Equation \ref{eq:IntegralEquation} has the same form as the volume integral equation governing the evolution of the classical polarization density field induced in the dielectric object by a classical source, that is, the classical electromagnetic scattering in the time domain (e.g., \cite{van_bladel_electromagnetic_2007}). It can be numerically solved using finite elements (e.g., \cite{van_bladel_electromagnetic_2007}), either by resorting to marching-on-time techniques (e.g., \cite{rao_time_1999}) or by frequency domain analysis and Fourier transform (for details, see Appendix E). In the frequency domain, the equation \ref{eq:IntegralEquation} can also be transformed into an equivalent surface integral equation that significantly reduces the computational burden (e.g., \cite{harrington_time-harmonic_2007}). The driving c-vector field ${\boldsymbol{D}}_\beta^{(\alpha)}$ depends on the index $\nu$, $\mu$ and $m$. However, as we shall show in Section VI only a few of the $\{\boldsymbol{p}_\beta^{(\alpha)}\}$ c-vector fields contribute to the evolution of statistical functions of the observables, depending on the initial quantum state of the system. 

\subsection{Expansion of the electric and magnetic field operators}

To evaluate the electric field operator, it is convenient to express it by grouping all the terms depending on the annihilation operators $\hat{a}_\mu$ and $\hat{c}_{m,\nu}$, and all the terms depending on the creation operators $\hat{a}_\mu^\dagger$ and $\hat{c}_{m,\nu}^\dagger$. We obtain the following.
\begin{equation}
  \hat{\mathbf{{E}}}\rt=   \hat{\mathbf{{E}}}^{(+)}\rt+ h.c.
\end{equation}
where $\hat{\mathbf{E}}^{(+)}$ contains only annihilation operators and the hermitian conjugate contains only creation operators. The term $\hat{\mathbf{E}}^{(+)}$, in turn, has two contributions: one from radiation and the other from matter. We express it as follows,
\begin{equation}
  \hat{\mathbf{{E}}}^{(+)}\rt=\hat{\mathbf{{E}}}^{(+)}_{rad}\rt+\hat{\mathbf{{E}}}^{(+)}_{mat}\rt,
\end{equation}
where
\begin{subequations}
\begin{align}
  \hat{\mathbf{{E}}}_{rad}^{(+)}\rt&=
  \sum_\mu \boldsymbol{E}_\mu^{(rad)}\rt\hat{{{{a}}}}_\mu,\\
  \hat{\mathbf{{E}}}_{mat}^{(+)}\rt&= \sum_{m,\nu} \boldsymbol{E}_{m,\nu}^{(mat)}\rt\hat{{{{c}}}}_{m,\nu}.
\end{align}
\end{subequations}
The c-fields $\boldsymbol{E}_\mu^{(rad)}\rt$ and $\boldsymbol{E}_\mu^{(mat)}\rt$ are given by 
\begin{subequations}
\begin{align}
\boldsymbol{E}_\mu^{(rad)}&=\boldsymbol{e}_{\mu}^{(rad)}\rt+\boldsymbol{e}_{\mu}\rt,\\
\boldsymbol{E}_{m,\nu}^{(mat)}&=\boldsymbol{e}_{{m,\nu}}^{(mat)}\rt+\sqrt{\frac{\hbar \sigma(\nu)}{\nu \pi}}\boldsymbol{N}_{{m}}\rt
\end{align}
\end{subequations}
where $\mathbf{e}_\mu$ is given by \ref{eq:e_mu}, $\boldsymbol{N}_{{m}}$ is given by \ref{eq:Efree_4}, and
\begin{equation}
  \boldsymbol{e}_\beta^{(\alpha)}\rt =\mathcal{L}\{\boldsymbol{p}_\beta^{(\alpha)}\}\rt,
\end{equation}
for $\alpha = rad$, $\beta =\mu$, and $\alpha = mat$, $\beta =m,\nu$. The c-vector fields $\boldsymbol{p}_\beta^{(\alpha)}$ and $\dot{\boldsymbol{p}}_\beta^{(\alpha)}$ are extended to $t<0$ in such a way as to be null.

Similarly, we proceed to evaluate the magnetic field operator. We express $\hat{\mathbf{{B}}}\rt$ as
\begin{equation}
  \hat{\mathbf{{B}}}\rt=   \hat{\mathbf{{B}}}^{(+)}\rt+  h.c.
\end{equation}
where
\begin{equation}
\hat{\mathbf{{B}}}^{(+)}\rt=\hat{\mathbf{{B}}}^{(+)}_{rad}\rt+\hat{\mathbf{{B}}}^{(+)}_{mat}\rt,
\end{equation}
and
\begin{subequations}
\begin{align}
  \hat{\mathbf{{B}}}_{rad}^{(+)}\rt&=
  \sum_\mu \boldsymbol{B}_\mu^{(rad)}\rt\hat{{{{a}}}}_\mu,\\
  \hat{\mathbf{{B}}}_{mat}^{(+)}\rt&= \sum_{m,\nu} \boldsymbol{B}_{m,\nu}^{(mat)}\rt\hat{{{{c}}}}_{m,\nu}.
\end{align}
\end{subequations}
The c-fields $\boldsymbol{B}_\mu^{(rad)}\rt$ and $\boldsymbol{B}_\mu^{(mat)}\rt$ are given by 
\begin{subequations}
\begin{align}
\boldsymbol{B}_\mu^{(rad)}&=\boldsymbol{b}_{\mu}^{(rad)}\rt+\boldsymbol{b}_{\mu}\rt,\\
\boldsymbol{B}_{m,\nu}^{(mat)}&=\boldsymbol{b}_{{m,\nu}}^{(mat)}\rt,
\end{align}
\end{subequations}
where $\mathbf{b}_\mu=\mathbf{k}\times\mathbf{e}_\mu/\omega_\mu$ and
\begin{equation}
  \boldsymbol{b}_\beta^{(\alpha)}\rt =\mu_0 \nabla \times \int_V\frac{\dot{\boldsymbol{{\mathbf{p}}}}_\beta^{(\alpha)}(\mathbf{r}';t')}{4\pi\left| {\bf r} - {\bf r}' \right|}\dV'.
\end{equation}
\\
\section{Statistical functions}
The knowledge of the c-vector fields $\boldsymbol{p}_\beta^\alpha \rt$ allows one to evaluate the statistical functions of any observable of the system, such as expectation values, uncertainty and correlation functions. In the following, as an example, we consider the \textit{single counting rate} for photoelectric measurements. 

The single counting rate $w_I$ is the mean value in the initial state $|\psi_0\rangle$ of the observable (e.g., \cite{cohen-tannoudji_photons_1997})
\begin{equation}
    \hat{I}\rt=\hat{\mathbf{E}}^{(-)}\rt\cdot\hat{\mathbf{E}}^{(+)}\rt,
\end{equation}
where $\mathbf{\hat{E}}^{(-)}\rpt=[\mathbf{\hat{E}}^{(+)}\rpt]^\dagger$; $\hat{I}\rt $ is arranged in the normal order (that is, all the annihilation operators on the right and all the creation operators on the left). Therefore, we have 
\begin{equation}
    w_I\rpt=\langle{\psi_0}|\hat{\mathbf{{E}}}^{(-)}\rt\cdot \hat{\mathbf{{E}}}^{(+)}\rt|\psi_0\rangle.
\end{equation}
We assume that
\begin{equation}
    |\psi_0\rangle=|\gamma_0\rangle \otimes |\phi_0\rangle
\end{equation}
where $\otimes$ denotes the tensor product, $|\gamma_0\rangle$ is the initial state of the matter field, and $|\phi_0\rangle$ is the initial state of the radiation field (they are supposed to be initially non-interacting). The matter field is initially supposed to be in thermodynamic equilibrium with a reservoir at temperature $T_0$. Instead, the radiation field is assumed to be in an eigenstate $|n_1, n_2, ..., n_\mu, ...\rangle$ of $\hat{H}_{rad}$, where $n_\mu$ is the number of photons in the mode $\mu$.
In this initial state, the contributions of the mixed terms $\hat{\mathbf{{E}}}_{mat}^{(-)}\cdot\hat{\mathbf{{E}}}^{(+)}_{rad}$ and $\hat{\mathbf{{E}}}_{rad}^{(-)}\cdot\hat{\mathbf{{E}}}_{mat}^{(+)}$ vanish, thus we obtain
\begin{equation}
    w_I\rpt= w_I^{(rad)}\rpt+w_I^{(mat)}\rpt
\end{equation}
where
\begin{multline}
    w_I^{(rad)}\rpt= \langle{\phi_0}|\hat{\mathbf{{E}}}_{rad}^{(-)}\cdot\hat{\mathbf{{E}}}^{(+)}_{rad}|\phi_0\rangle= \\ \sum_\mu \sum_{\mu'} [\mathbf{E}_{\mu}^{(rad)}\rpt]^* \cdot \mathbf{E}_{\mu'}^{(rad)}\rpt\langle{\phi_0}|\hat{{{{a}}}}_{\mu}^\dagger \hat{{{{a}}}}_{\mu'}|\phi_0\rangle,
\end{multline}
and
\begin{multline}
    w_I^{(mat)}\rpt= \langle{\gamma_0}|\hat{\mathbf{{E}}}_{mat}^{(-)}\cdot\hat{\mathbf{{E}}}^{(+)}_{mat}|\gamma_0\rangle=\\ \sum_{m,\nu} \sum_{m',\nu'} [\mathbf{E}_{m,\nu}^{(mat)}\rpt]^* \cdot \mathbf{E}_{m',\nu'}^{(mat)}\rpt\langle{\gamma_0}|\hat{{{{c}}}}_{m,\nu}^\dagger \hat{{{{c}}}}_{m',\nu'}|\gamma_0\rangle.
\end{multline}

To further simplify the discussion, we assume that the radiation field is initially in the one-photon state (e.g., \cite{cohen-tannoudji_photons_1997})
\begin{equation}
|\phi_0\rangle=\sum_{\mu \in \mathcal{M}_0} b_\mu \hat{a}_\mu^\dagger|0\rangle_r
\end{equation}
and the free polarization operator is uniformly distributed in the dielectric object. The ket $|0\rangle_r$ is the vacuum state for the radiation field, $b_\mu$ is the mode amplitude and $\mathcal{M}_0$ denotes the set of transverse plane wave modes that are initially in the one-photon state. The creation operator $\hat{a}_\mu^\dagger$ acting on the vacuum state $|0\rangle_r$ gives a state with one photon $\mathbf{k}$. The mode amplitude $b_\mu$ satisfies the normalization condition $\sum_{\mu \in \mathcal{M}_0}|b_\mu|^2=1$. 
The contribution of the initial state of the reservoir to the counting rate is given by
\begin{widetext}
\begin{equation}
    w_I^{(mat)}\rpt=\int_0^\infty d\nu \int_0^\infty d\nu' [\mathbf{E}_{\nu}^{(mat)}\rpt]^* \cdot \mathbf{E}_{\nu'}^{(mat)}\rpt\langle{\gamma_0}|\hat{{{{c}}}}_{\nu}^\dagger \hat{{{{c}}}}_{\nu'}|\gamma_0\rangle
\end{equation}
\end{widetext}
where $\mathbf{E}_{\nu}^{(mat)}\rpt$ is the c-electric field generated by a uniform distribution of polarization in the dielectric object,
\begin{equation}
   \langle{\gamma_0}|\hat{{{{c}}}}_{\nu}^\dagger \hat{{{{c}}}}_{\nu'}|\gamma_0\rangle = \rho_\nu \delta(\nu-\nu')
\end{equation}
and
\begin{equation}
   \rho_\nu = \frac{1}{e^{\hbar\nu/k_B T_0}-1}
\end{equation}
(reservoir oscillators with different frequencies are initially uncorrelated). Therefore, we obtain the following
\begin{equation}
    w_I^{(mat)}\rpt=\int_0^\infty d\nu  \rho_\nu|\mathbf{E}_{\nu}^{(mat)}\rpt|^2.
\end{equation}
The contribution of the initial state of the radiation field to the counting rate is given by
\begin{equation}
    w_I^{(rad)}\rpt=|\sum_{\mu \in \mathcal{M}_0} b_\mu\mathbf{E}_{\mu}^{(rad)}\rpt|^2.
\end{equation}
In the scenario considered, only the wave modes with $\mu \in \mathcal{M}_0$ contribute to $w_I^{(rad)}$. 
\\
\section{Discussion}

In this paper, we have considered the problem of the interaction between a finite-size dispersive dielectric object and the electromagnetic field in the framework of macroscopic quantum electrodynamics using a Hopfield-type approach. We have proposed an integral equation for the polarization density field operator in the Heisenberg picture based on the retarded Green function for the vacuum. We have expressed the electric and magnetic field operators as functions of the polarization density field operator through the same retarded Green function. The known term of the integral equation is a linear combination of the annihilation and creation operators of the bosonic matter field and of the radiation field in the Schrodinger picture, through which the initial quantum state of the system comes into play.
 
We have used the Coulomb gauge to derive the equation (\ref{eq:Pol}-\ref{eq:Boper}) because the use of the Lorenz gauge would have been much more challenging. The transverse component of the electric field operator $-{\dot{\hat{\mathbf{\A}}}}$ depends on the polarization current density field operator ${\dot{\hat{\mathbf{P}}}}$ through the transverse dyadic Greeen function for the vacuum. The longitudinal component of the electric field operator ${\mathbf{E}}^\parallel\{\hat{\mathbf{P}}\}$ depends on the polarization charge density operator through the static dyadic Green function for the vacuum. The longitudinal and
transverse components of the electric field operators
have no separate physical meaning; only their sum is physical. As in classical electrodynamics (e.g., \cite{jackson_historical_2001}, \cite{jean_g_van_bladel_singular_1996}), summing $-{\dot{\hat{\mathbf{\A}}}}$ and ${\mathbf{E}}^\parallel\{\hat{\mathbf{P}}\}$ and using standard mathematical manipulations, we have obtained the electric field operator as a function of the polarization field density operator in terms of the retarded Green function for the vacuum.

 Within the Lorenz gauge, the vector and scalar potentials are treated on the same foot, but the quantization is not as straightforward as in the Coulomb gauge (e.g., \cite{w_heitler_quantum_1984}, \cite{cohen-tannoudji_photons_1997}). The quantization in the Lorenz gauge does not require the transverse delta function, because the two transverse photons (associated with the transverse components of the vector potential), the longitudinal photon (associated with the longitudinal component of the vector potential), and the scalar photon (associated with the scalar potential) are considered as independent. The Lorenz gauge is enforced as a subsidiary condition after quantization.
This constraint cannot be imposed as an operator identity, because the vector and scalar potential operators act on different subspaces of the state space of the system, e.g. \cite{cohen-tannoudji_photons_1997}. Furthermore, the Lorenz gauge is not compatible with the commutation relations between the electromagnetic potential operators and the corresponding conjugate momentum operators, e.g. \cite{w_heitler_quantum_1984}.
The Lorenz gauge can be enforced only for the expectation values and not for the operators: this constraint selects the possible ``physical states'' \cite{cohen-tannoudji_photons_1997}.
The commutation relation for the creation and annihilation operators of the scalar photons has a minus sign compared with the commutation relation of the transverse and longitudinal photons.
As a consequence, there exist states with a negative norm \cite{cohen-tannoudji_photons_1997}, which is wholly unacceptable in a Hilbert space, where all norms must be positive to allow a probabilistic interpretation. Furthermore, the contribution of the scalar potential operator to the Hamiltonian is negative \cite{cohen-tannoudji_photons_1997}. In the classical framework, the energy associated with the longitudinal component of the vector potential compensates exactly that associated with the scalar potential because of the Lorenz condition and the overall Hamiltonian is positive, while in the quantum framework this does not happen. To overcome these difficulties, a covariant quantization with an ``indefinite metric'' is required (for more details, see, for example, in \cite{cohen-tannoudji_photons_1997} and \cite{w_heitler_quantum_1984}). Cohen-Tannoudji et al. demonstrated the transition from the Lorentz to the Coulomb gauge by applying a  unitary transformation with respect to the indefinite metric together with the subsidiary condition  [ see Complement $\S B_\text{V}.1 $ d) of Ref. \cite{cohen-tannoudji_photons_1997} ]

\section{Conclusions}
 
The principal outcomes of our integral formulation are the following. It enables the direct application of the consolidated repertoire in computational classical electrodynamics to evaluate the time evolution of the observables of the system and the statistical functions. The unknown and known terms of Eq. \ref{eq:Pol} are operators: the known term includes the effect of fluctuations of the bosonic matter field and of the radiation field. Since the known term is a linear combination of creation and annihilation operators for the matter and the radiation fields, and Eq. \ref{eq:Pol} is linear, we expand the polarization density field operator in terms of creation and annihilation operators and then we obtain the set of integral equations \ref{eq:IntegralEquation} where the unknowns are c-vector fields. These equations have the same structure as the volume integral equation we encounter in classical electromagnetic scattering (e.g., \cite{van_bladel_electromagnetic_2007}). They can be solved numerically by finite elements, either by resorting to marching-on-time techniques or by frequency domain analysis and Fourier transform. In the same manner, we evaluate the expressions \ref{eq:Eoper} and \ref{eq:Boper}.

The proposed formulation is very general:

i) In the presence of a quantum emitter, in Eq. \ref{eq:Pol} we have to add the contribution due to the electric field generated by the electric dipole moment operator of the quantum emitter, which can be expressed using the operator \ref{eq:OperatorL}. In this case, we have also to add the Heisenberg equation for the electric dipole moment operator of the quantum emitter. 

ii) In the presence of a classical source, in the expression of $\hat{\mathbf{D}}_f$ we have to add the contribution due to the electric field generated by the classical source using again the operator \ref{eq:OperatorL}.

iii) In presence of an inhomogeneous dielectric, the coupling parameter between the polarization density field and the electric field, $\alpha_\nu$, depends on the points of space. The polarization density field is no longer solenoidal in $V$. The Coulomb interaction energy term ${H}_{Coul}$ would also contain an additional term due to the volumetric polarization charge density $-\nabla \cdot \boldsymbol{P}$, in addition to $P_{n}$. However, equations \ref{eq:Pol} and \ref{eq:Ftot}-\ref{eq:Boper} continue to hold as long as: a) the term $\nabla \left[ \frac{1}{4 \pi \varepsilon_0} \int_{V} \,\frac{{{\nabla \cdot\hat{\mathbf{P}}}}( \mathbf{r}';t')}{\left| {\bf r} - {\bf r}' \right|} \dV'\right]$ is added to the expression of $\mathcal{L}\{\hat{\mathbf{P}}\}\rt$ given by \ref{eq:OperatorL};
b) an analogous term is added to the expression of $\boldsymbol{N}_m(\mathbf{r};t)$ given by \ref{eq:Efree_4}, providing that the orthonormal real basis $\{\mathbf{U}_m(\mathbf{r})\}$ can also represent nonsolenoidal vector fields.

This paper enables the application of classical computational electromagnetic methods, including boundary integral methods and multilevel fast multipoles, to the modeling of the dynamics of realistic quantum devices.
\\
\begin{acknowledgments}
Giovanni Miano acknowledges financial support from PNRR MUR project CN00000013-ICSC, Carlo Forestiere acknowledges financial support from PNRR MUR project PE0000023-NQSTI.
\end{acknowledgments}

\appendix
\section{Longitudinal and transverse dyadic delta functions}
\label{sec:longitudinal}

A longitudinal vector field in $V_\infty$ is everywhere irrotational (its tangential component to any surface is continuous), while a transverse vector field in $V_\infty$ is everywhere solenoidal (its normal component to any surface is continuous), e.g., \cite{cohen-tannoudji_photons_1997}, \cite{arnoldus_transverse_2003}, \cite{jean_g_van_bladel_singular_1996}. The longitudinal component $\boldsymbol{C}^\parallel(\rb)$ of a regular vector field $\boldsymbol{C}(\rb)$ defined in $V_\infty$ is given by the spatial convolution between the longitudinal dyadic delta function
\begin{equation}
\overleftrightarrow{\boldsymbol\delta}^\parallel(\rb)=-\boldsymbol\nabla\boldsymbol\nabla(1/4\pi r)
\end{equation}
and $\boldsymbol{C}(\rb)$. Instead, the transverse component $\boldsymbol{C}^\perp(\rb)$ is given by the spatial convolution with the transverse dyadic delta function
\begin{equation}
    \overleftrightarrow{\boldsymbol\delta}^\perp(\rb)=\I\delta(\rb)- \overleftrightarrow{\boldsymbol\delta}^\parallel(\rb)
\end{equation}
where $\I$ is the three-dimensional unit tensor, and $\delta(\bf r)$ is the Dirac delta function in the three dimensional space. 
The distribution $\boldsymbol\nabla\boldsymbol\nabla(1/4\pi r)$ can be expressed as (e.g., \cite{belinfante_longitudinal_1946}, \cite{jean_g_van_bladel_singular_1996}) 
\begin{equation}
\label{eq:Princ}
\boldsymbol\nabla\boldsymbol\nabla(1/4\pi r)=-\frac{\boldsymbol\delta(\boldsymbol{r})}{3}-PV_{S}\left(\frac{\overleftrightarrow{I}-3\mathbf{r}_u{\mathbf{r}}_u}{4\pi r^3}\right)
\end{equation}
where $\mathbf{r}_u=\mathbf{r}/r$ and $PV_{S}$ denotes the principal value based on a spherical cavity centered at $\boldsymbol{r}=\boldsymbol{0}$.

The vector fields $\boldsymbol{C}^\parallel$ and $\boldsymbol{C}^\perp$ are orthogonal according to the scalar product $\langle {\boldsymbol{F}}, {\bf G} \rangle = \int_{V_\infty}{\boldsymbol{F}}^* \rp \cdot \boldsymbol{G} \rp \dV$.

\section{Solution of Eq. \ref{eq:Adotdot}}
\label{sec:vectpot}

In this appendix, we solve equation \ref{eq:Adotdot}
with the initial conditions \ref{eq:inita} and \ref{eq:initdotA}, and we express the vector potential operator as a function of the polarization density field operator. 

\subsection{Vector potential operator}
To determine the solution of equation \ref{eq:Adotdot} we represent the vector potential operator in terms of transverse plane waves in free space \ref{eq:planewave},
\begin{equation}
\label{eq:serie1}
    \hat{\Ab}\rt = \sum_{\mu} \hat{A}_\mu(t) \mathbf{w}_\mu \rp,
\end{equation}
where $\{\hat{A}_{\mu}(t)\}$ are the coordinate operators of the vector potential. Substituting \ref{eq:serie1} into equation \ref{eq:Adotdot} and projecting both sides of the equation onto the transverse plane waves we obtain for any $\mu \in \mathcal{M}$:
\begin{equation}
    \label{eq:rad}
    \ddot{\hat{A}}_\mu+\omega_\mu^2\hat{A}_\mu=\frac{1}{\varepsilon_0}\langle \mathbf{w}_\mu,\dot{\hat{\mathbf{P}}}\rangle.
\end{equation}
We have used the property $\langle \mathbf{w}_\mu,\dot{\hat{\mathbf{P}}}^\perp\rangle=\langle \mathbf{w}_\mu,\dot{\hat{\mathbf{P}}}\rangle$.
Solving equation \ref{eq:rad} we obtain
\begin{equation}
\label{eq:dotA}
    {\hat{A}}_\mu = \frac{1}{\varepsilon_0} g_{\mu}(t)* \langle \mathbf{w}_\mu,\dot{\hat{\mathbf{P}}}\rangle+ \hat{\alpha}_\mu(t)
\end{equation}
where
\begin{equation}
   g_{\mu}\tp = \frac{1}{\omega_\mu} u \tp \sin(\omega_\mu t),
\end{equation}
$u \tp$ is the Heaviside function,
\begin{equation}
\label{eq:E0}
   \hat{\alpha}_\mu(t) =\hat{A}_{\mu}^{(S)} \cos(\omega_\mu t) + \frac{1}{\varepsilon_0 \omega_\mu}{\hat{\Pi}}_\mu^{(S)} \sin(\omega_\mu t),
\end{equation}
$\hat{A}_{\mu}^{(S)} = \langle \mathbf{w}_{\mu},\hat{\mathbf{A}}^{(S)}\rangle$ and $\hat{\Pi}_{\mu}^{(S)} = \langle \mathbf{w}_{\mu},\hat{\mathbf{\Pi}}^{(S)}\rangle$.
Therefore, the vector potential operator is given by
\begin{equation}
\label{eq:dotA3}
{\hat{\Ab}}\rt = \frac{1}{\varepsilon_0}\sum_{\mu} [g_\mu (t)*\langle \mathbf{w}_\mu,\dot{\hat{\mathbf{P}}}\rangle] \mathbf{w}_\mu \rp + \hat{\mathbf{A}}_f\rt ,
\end{equation}
where
\begin{equation}
\Af \rt =\sum_{\mu} \hat{\alpha}_\mu(t) \mathbf{w}_\mu \rp
\end{equation}
takes into account the contributions of the initial conditions of the radiation field operators. 

\subsection{Transverse dyadic Green function}

Using the identity
\begin{multline}
\sum_{\mu} [g_\mu (t)*\langle \mathbf{w}_\mu,\dot{\hat{\mathbf{P}}}\rangle] \mathbf{w}_\mu \rp = \\ \int_V \dV' \left[\sum_{\mu} g_\mu (t)\mathbf{w}_\mu(\mathbf{r})\mathbf{w}_\mu^*(\mathbf{r}')\right]*\dot{\hat{\mathbf{P}}}(\mathbf{r}';t),
\end{multline}
we obtain from the relation \ref{eq:dotA3}
\begin{multline}
    \hat{\mathbf{A}}\rt =\mu_0 \int_V\dV'\int_0^\infty dt' \overleftrightarrow{g}^\perp \left(\mathbf{r}- \mathbf{r}';t-t'\right) \dot{\hat{\mathbf{P}}}(\mathbf{r}';t') \\ + \Af \rt
\end{multline}
where the transverse dyadic Green function $\overleftrightarrow{g}^\perp \left(\mathbf{r}- \mathbf{r}';t\right)$ is given by
\begin{equation}
\label{eq:serie}
   \overleftrightarrow{g}^\perp \left(\mathbf{r}- \mathbf{r}';t\right) = c_0^2 \sum_{\mu} g_\mu (t)\mathbf{w}_\mu(\mathbf{r})\mathbf{w}_\mu^*(\mathbf{r}').
\end{equation}

We now use the Laplace transform to evaluate \ref{eq:serie}. The Laplace transform of $\overleftrightarrow{g}^\perp \left(\mathbf{r}- \mathbf{r}';t\right)$ is
\begin{equation}
    \overleftrightarrow{G}^\perp(\rb-\rb';s)=\sum_{\mu}\frac{1}{k^2+s^2/c_0^2} \mathbf w_{\mu}(\rb)\,\mathbf w_{\mu}^*(\rb').
\end{equation}
Using the expression of $\mathbf w_{q}(\rb)$ (see \ref{eq:planewave}) we obtain the following
\begin{equation}
\label{eq:ExpS2}
    \overleftrightarrow{G}^\perp(\rb -\rb';s)=\frac{1}{(2\pi)^3}\int{} d^3\mathbf{k} \,\overleftrightarrow{\mathcal{G}}^\perp(\mathbf{k};s)\,e^{i\mathbf{k}\cdot(\rb-\rb')} 
\end{equation}
where
\begin{equation}
\label{eq:FourierI}
    \overleftrightarrow{\mathcal{G}}^\perp(\mathbf{k};s)= \frac{1}{k^2+s^2/c_0^2} (\overleftrightarrow{I}-\hat{\mathbf{k}}\,\hat{\mathbf{k}})
\end{equation}
is the transverse dyadic Green function for the vector potential in the wavenumber domain and in free space. By evaluating the Fourier integral \ref{eq:ExpS2} we obtain \cite{arnoldus_transverse_2003}
\begin{multline}
    \overleftrightarrow{G}^\perp \left(\rb;s \right)=\frac{e^{-sr/c_0}}{4\pi r}(\overleftrightarrow{I}-{\mathbf{r}_u}{\mathbf{r}_u})+\\ (\overleftrightarrow{I}-3\mathbf{r}_u\mathbf{r}_u)\frac{c_0}{4\pi sr^2}\left[e^{-sr/c_0}-\frac{c_0}{sr}(1-e^{-sr/c_0})\right]
\end{multline}
where $\mathbf{r}_u=\mathbf{r}/r$. This expression can be rewritten as
\begin{equation}
\label{eq:GreenT}
    \overleftrightarrow{G}^\perp \left(\rb;s \right)=\overleftrightarrow{G} \left(\rb;s \right) - \overleftrightarrow{G}^\parallel \left(\rb;s \right),
\end{equation}
where 
\begin{multline}
\overleftrightarrow{G} \left(\rb;s \right) =\frac{c_0^2}{3s^2}\boldsymbol\delta(\boldsymbol{r})+ \frac{e^{-sr/c_0}}{4\pi r}\times \\
\left[(\overleftrightarrow{I}-{\mathbf{r}}_u{\mathbf{r}}_u)
  + \frac{c_0}{sr}{(\overleftrightarrow{I}-3\mathbf{r}_u{\mathbf{r}}_u)}(1+\frac{c_0}{sr})\right]
\end{multline}
and
\begin{equation}
\label{eq:Green_0}
  \overleftrightarrow{G}^\parallel \left(\rb;s \right)= \frac{c_0^2}{s^2}\left[\frac{\boldsymbol\delta(\boldsymbol{r})}{3}+\frac{(\overleftrightarrow{I}-3\mathbf{r}_u{\mathbf{r}}_u)}{4\pi r^3}\right].
\end{equation}
The dyad $\overleftrightarrow{G}\left(\rb;s \right)$ is the Green function for the vector potential in the temporal gauge, $\overleftrightarrow{G}^\perp \left(\rb;s \right)$ is its transverse component, and $\overleftrightarrow{G}^\parallel \left(\rb;s \right)$ is its longitudinal component. We observe that $\overleftrightarrow{G}^\perp \left(\rb;s \right)$ does not contain the Dirac delta function $\boldsymbol\delta(\boldsymbol{r})$. 

The dyad $\overleftrightarrow{G}\left(\rb;s \right)$is the Green function in vacuum for the retarded electric field. It can be expressed through the scalar Green function for the vacuum as (e.g., \cite{jean_g_van_bladel_singular_1996})
\begin{equation}
\label{eq:Green10}
    \overleftrightarrow{G}(\rb;s)=G(\rb;s)\overleftrightarrow{I}-\frac{c_0^2}{s^2}\nabla\nabla G(\rb;s)
\end{equation}
where
\begin{equation}
 {G}(\rb;s) = \frac{e^{-sr/c_0}}{4\pi r}.
\end{equation}
The expression $\nabla\nabla G(\rb;s)$ is a distribution to be evaluated according to \ref{eq:Princ}. The longitudinal component $\overleftrightarrow{G}^\parallel \left(\rb;s \right)$ can be also expressed through the scalar Green function $G$,
\begin{equation}
\label{eq:Green_01}
  \overleftrightarrow{G}^\parallel \left(\rb;s \right)=-\frac{c_0^2}{s^2}\nabla\nabla G(\boldsymbol{r};s=0).
\end{equation}
Combining \ref{eq:GreenT}, \ref{eq:Green10} and \ref{eq:Green_01} we obtain
\begin{equation}
\label{eq:GreenT10}
     \overleftrightarrow{G}^\perp \left(\rb;s \right)=G(\rb;s)\overleftrightarrow{I}-\frac{c_0^2}{s^2}\nabla \nabla [G(\rb;s)-G(\rb;s=0)].
\end{equation}

In the time domain, expressions \ref{eq:GreenT}-\ref{eq:Green_0} become
\begin{equation}
\label{eq:GreenTravTime0} 
    \overleftrightarrow{g}^\perp \left(\rb;t \right)=\overleftrightarrow{g} \left(\rb;t \right) - \overleftrightarrow{g}^\parallel \left(\rb;t \right), \quad
\end{equation}
where
\begin{multline}
\label{eq:GreenTravTime1} 
    \overleftrightarrow{g} \left(\rb;t \right)=c_0^2\frac{\boldsymbol\delta(\boldsymbol{r})}{3}u(t)t+\frac{\overleftrightarrow{I}-{\mathbf{r}_u}{\mathbf{r}_u}}{4\pi r}\delta(t-r/c_0)+ \\
    c_0\frac{\overleftrightarrow{I}-3{\mathbf{r}_u}{\mathbf{r}_u}}{4\pi r^2}\,u(t-r/c_0)\left[1+\frac{c_0}{r}\,(t-r/c_0))\right],
\end{multline}
$\mathbf{r}_u=\mathbf{r}/r$, $u(t)$ is the Heaviside function, and
\begin{multline}
\label{eq:GreenTravTime2} 
    \overleftrightarrow{g}^\parallel\left(\rb;t \right)=\left[\frac{\boldsymbol\delta(\boldsymbol{r})}{3}+\frac{(\overleftrightarrow{I}-3\mathbf{r}_u{\mathbf{r}}_u)}{4\pi r^3}\right] c_0^2u(t)t=\\
    -\nabla\nabla\left(\frac{1}{4\pi r}\right)c_0^2u(t)t.
\end{multline}

\subsection{Free vector potential field operator}

The vector field operator $\hat{\mathbf{A}}_f$ takes into account the contribution of the initial conditions of the radiation field operators. We call it \textit{free vector potential field operator}. The operators $\{\hat{A}_{\mu}^{(S)}\}$ and the operators $\{\hat{\Pi}_{\mu}^{(S)}\}$ obey the commutation relations
\begin{equation}
\label{eq:etcra}
\left[ \hat{\Pi}_{\mu'}^{(S)}, \hat{A}_{\mu}^{\dagger (S)}\right] = -i \delta_{s,s'} \delta \left( \mathbf{k} - \mathbf{k}' \right),
\end{equation}
for any couple $\mu, \mu'\in \mathcal{M}$, while all other commutators vanish. We express them in terms of the annihilation and creation operators  
 $\hat{a}_\mu$ and $\hat{a}^\dagger_\mu$ for the transverse electromagnetic field mode $\mu$, in the Schr\"odinger picture, defined as:
 \begin{equation}
     \hat{a}_\mu = \frac{\omega_\mu}{2 \mathcal{E}_\mu} \left( \hat{A}^{(S)}_\mu + \frac{i}{\omega_\mu} \hat{\Pi}_\mu^{(S)} \right)
 \end{equation}
 where $\mathcal{E}_\mu=\{\hbar \omega_\mu/[2\varepsilon_0 (2\pi)^3]\}^{1/2}$. They obey the commutation relation $[\hat{a}_\mu,\hat{a}_{\mu'}^\dagger]=\delta(\mu-\mu'),\,  [\hat{a}_\mu,\hat{a}_{\mu'}]=0,\, [\hat{a}_\mu^\dagger,\hat{a}_{\mu'}^\dagger]=0$.
The known vector field operator $\Af\rpt$, then, is rewritten as
\begin{equation}
\label{eq:Afree3}
\Af\rpt =\sum_\mu\frac{1}{i\omega_\mu}\boldsymbol{e}_\mu\rt \hat{a}_\mu+h.c.,
\end{equation}
where
\begin{equation}
    {\boldsymbol{e}}_\mu\rpt =i\mathcal{E}_\mu \mathbf{w}_\mu\rp e^{-i\omega_\mu t}.
\end{equation}

\section{Solution of Eq. \ref{eq:Ydotdot}}
\label{sec:susceptibility}

Here, we first solve the equation \ref{eq:Ydotdot}
with the initial conditions \ref{eq:inity1} and \ref{eq:inity2}, where $\hat{\mathbf{E}}=-\dot{\hat{\mathbf{A}}}+\mathbf{E}^\parallel\{\hat{\mathbf{P}}\}$. Then, we evaluate the relation between the polarization density field operator and the electric field operator. 

\subsection{Polarization density field operator}
In the Laplace domain, Eq. \ref{eq:Ydotdot} becomes
\begin{equation}
\label{eq:YdotdotL}
(s^2+\nu^2) \hat{\boldsymbol{{\mathcal{Y}}}}_\nu=\alpha_\nu \hat{\boldsymbol{\mathcal{E}}} + [s\hat{\mathbf{Y}}_\nu^{(S)}(\mathbf{r})+\dot{\hat{\mathbf{Y}}}_\nu^{(S)}(\mathbf{r})]
\end{equation}
where $\hat{\boldsymbol{{\mathcal{Y}}}}_\nu\rs$ is the Laplace transform of ${\hat{\mathbf{Y}}}_\nu\rt$ and $\hat{\boldsymbol{{\mathcal{E}}}}\rs$ is the Laplace transform of ${\hat{\mathbf{E}}}\rt$. Therefore, the polarization density field operator is given by
\begin{equation}
\label{eq:PolAs}
 \hat{\boldsymbol{{\mathcal{P}}}}=\varepsilon_0{\Tilde{\chi}}(s) \hat{\boldsymbol{{\mathcal{E}}}}+ \hat{\boldsymbol{{\mathcal{P}}}}_f \quad \text{in}\, V
\end{equation}
where
\begin{equation}
\label{eq:chiTilde}
{\Tilde{\chi}}(s)=\frac{1}{\varepsilon_0}\int_0^\infty d\nu\, \frac{\alpha_\nu^2}{s^2+\nu^2},
\end{equation}
and
\begin{equation}
\label{eq:P0}
 \hat{\boldsymbol{{\mathcal{P}}}}_f (\mathbf{r};s)=\int_0^\infty d\nu\, \frac{\alpha_\nu}{s^2+\nu^2} [s{\hat{\mathbf{Y}}}_\nu^{(S)}(\mathbf{r})+\dot{\hat{\mathbf{Y}}}_\nu^{(S)}(\mathbf{r})].
\end{equation}
Due to dissipation, the region of convergence of the Laplace transform contains the imaginary axis; therefore, we evaluate ${\Tilde{\chi}}(s)$ for $s=i\omega + \epsilon$ where $\epsilon \downarrow 0$.
By using the relation (e.g., \cite{w_heitler_quantum_1984})
\begin{equation}
\label{eq:Princv}
  \frac{1}{x-i\epsilon}= i\pi\delta(x) + \mathcal{P}\frac{1}{x},
\end{equation}
where $\mathcal{P}$ denotes the Cauchy principal value, we obtain the susceptibility of the object in the frequency domain $\chi(\omega)= {\Tilde{\chi}}(s=i\omega+\epsilon)$
\begin{equation}
  \varepsilon_0\chi(\omega)= \mathcal{P}\int_0^\infty d\nu\, \frac{\alpha_\nu^2}{\nu^2-\omega^2} -i\frac{\pi}{2}\frac{\alpha_\omega^2}{\omega}.
\end{equation}
Therefore, the coupling coefficient $\alpha_\nu$ is related to the imaginary part of the susceptibility $\chi_i$ through the relation
\begin{equation}
  \alpha_\nu=\sqrt{\frac{2\sigma(\nu)}{\pi}}
\end{equation}
where $\sigma(\nu) = -\varepsilon_0 \nu{\chi}_i(\nu)$.

In the time domain, the relation \ref{eq:PolAs} becomes
\begin{equation}
   \label{eq:Ptime}
    \hat{\mathbf{P}}\rpt = 
        \varepsilon_0 h_\chi (t) *\hat{\mathbf{E}} \left(\mathbf{r};t \right) + \Pf (\mathbf{r;t}) \quad \text{in}\, V
\end{equation}
where $h_\chi(t)$ is the inverse Fourier transform of $\chi(\omega)$,
\begin{equation}
    \Pf \rpt = \int_0^\infty d\nu\, \sqrt{\frac{2\sigma(\nu)}{\pi}} \hat{\mathbf{Y}}_\nu^{free}\rpt,
\end{equation}
and
\begin{equation}
   \hat{\mathbf{Y}}_\nu^{free}\rpt=\hat{\mathbf{Y}}_\nu^{(S)}(\mathbf{r}) \cos(\nu t) + \frac{1}{\nu}\dot{\hat{\mathbf{Y}}}_\nu^{(S)}(\mathbf{r}) \sin(\nu t).
\end{equation}
Applying the initial value theorem to the expression \ref{eq:chiTilde} we find $h_\chi (0)=0$.

Starting from Eq. \ref{eq:Ptime} we obtain immediately Eq. \ref{eq:PolA1bis} where $h_\eta (t)$ is the inverse Fourier transform of $\eta(\omega)=1/\chi(\omega)$.

\subsection{Free polarization density field operator}

The known vector field operator $\hat{\mathbf{P}}_f$ takes into account the contribution of the initial conditions of the matter field operators and of the vector potential operators to the polarization dynamics. We call it \textit{free  polarization density field operator}.
To represent $\Pf \rpt$ we introduce the discrete, orthonormal, real basis $\{\mathbf{U}_m(\mathbf{r})\}$ defined in $V$. We denote by $\{\hat{{y}}_{m,\nu}^{(S)}\}$ the coordinate operators of the matter field operator $\hat{\mathbf{Y}}_\nu^{(S)}$ and by $\{\hat{{q}}_{m,\nu}^{(S)}\}$ the coordinate operators of the conjugate field operator $\hat{\mathbf{Q}}_\nu^{(S)}$, that is, $\hat{\mathbf{Y}}_\nu^{(S)}\rp=\sum_m \mathbf{U}_m\rp \hat{y}_{m,\, \nu}^{(S)}$ and $\hat{\mathbf{Q}}_\nu^{(S)}\rp=\sum_m \mathbf{U}_m\rp \hat{q}_{m,\, \nu}^{(S)}$. The coordinate operators obey the commutation relations 
\begin{align}
\label{eq:etcryl}
\left[ \hat{q}_{m, \, \nu}^{(S)}, \hat{y}_{m',\, \nu'}^{(S)} \right] &= -i \hbar \delta\left(\nu-\nu'\right)\delta_{m m'},
\end{align}
 while all other commutators vanish. We represent the coordinate operators through the annihilation $\hat{{c}}_{m,\nu}$ and creation $\hat{{c}}_{m,\nu}^\dagger$ operators for the matter field in the Schr\"odinger picture. We have
\begin{equation}
\hat{{c}}_{m,\nu} = \sqrt{\frac{\nu}{2\hbar}} \left(\hat{{y}}_{m,\nu}^{(S)} + \frac{i}{\nu}{\hat{{q}}}_{m,\nu}^{(S)}\right).
\end{equation}
The annihilation and creation operators obey the commutation relations $[\hat{c}_{m,\,\nu},\hat{c}_{m',\,\nu'}^\dagger]=\delta(\mu-\mu')\delta(\nu-\nu'),\,  [\hat{c}_{m,\,\nu},\hat{c}_{m',\,\nu'}]=0$, and $[\hat{c}_{m,\,\nu}^\dagger,\hat{c}_{m',\,\nu'}^\dagger]=0$. Therefore, $\Pf \rpt$ is expressed as
\begin{equation}
\label{eq:Pfree1}
\Pf\rt = \mathbf{\hat{{P}}}_0\rt - \varepsilon_0 h_\chi (t)\hat{\mathbf{A}}^{(S)}(\mathbf{r}) \quad \text{in}\, V
\end{equation}
where
\begin{equation}
\mathbf{\hat{{P}}}_0\rt = \sum_{m,\,\nu} \sqrt{\frac{\hbar \sigma(\nu)}{\nu \pi}} \mathbf{U}_m(\mathbf{r})\,e^{-i\nu t}\hat{{c}}_{m,\nu} + h.c. ,
\end{equation}
 and $\sum_{m, \, \nu} \left( \cdot \right)= \displaystyle\sum_{m} \int_0^\infty d\nu \,\left( \cdot \right) $.

\section{Integral expressions for the electric and magnetic field operators}
\label{sec:Covarinat}
In this appendix, we obtain expressions \ref{eq:Eoper} and \ref{eq:Boper}.

\subsection{Electric field operator}
The expression of the electric field operator is given by $\hat{\mathbf{E}}=-\dot{\hat{\mathbf{A}}}+\mathbf{E}^\parallel\{\hat{\mathbf{P}}\}$. The longitudinal $\mathbf{E}^\parallel\{\hat{\mathbf{P}}\}$ and
transverse components $-\dot{\hat{\mathbf{A}}}$
have no separate physical meaning; only their sum is physical.
Using \ref{eq:Atime} and \ref{eq:Elong} we obtain
\begin{multline}
\label{eq:Etime10}
    \hat{\mathbf{E}}\rt =-\mu_0\frac{\partial}{\partial t} \int_V\dV'\int_0^t d\tau{{\overleftrightarrow{g}}}^\perp \left(\mathbf{r} - \mathbf{r}';t-\tau\right) \dot{\hat{\mathbf{P}}}(\mathbf{r}';\tau) \\ -\frac{1}{\varepsilon_0} \nabla\int_{\partial V} \,\frac{ \hat{\mathbf{P}} \left( \rb ';t\right)\cdot\,\mathbf{n}(\mathbf{r'})}{4\pi\left| {\bf r} - {\bf r}' \right|}\dS'+ \hat{\mathbf{E}}^\perp_f\rt
\end{multline}
where $\hat{\mathbf{E}}^\perp_f=-\dot{\hat{\mathbf{A}}}_f$. We now rewrite this expression in terms of the retarded Green function for the vacuum, as in classical covariant electrodynamics.

Let us consider the first term on the right hand side of \ref{eq:Etime10}, which we denote by $\int_V \hat{\mathbf{c}}(t; \rb,\rb')\dV'$ where
\begin{equation}
\hat{\mathbf{c}}(t; \rb,\rb')=-\mu_0\frac{\partial}{\partial t}\int_0^t {{\overleftrightarrow{g}}}^\perp \left(\mathbf{r} - \mathbf{r}';t-\tau\right) \dot{\hat{\mathbf{P}}}(\mathbf{r}';\tau)d\tau.
\end{equation}
The Laplace transform of $\hat{\mathbf{c}}(t; \rb,\rb')$ is
\begin{equation}
\label{eq:temp}
{\hat{\boldsymbol{C}}}(s;\rb,\rb')=-\mu_0s {{\overleftrightarrow{G}}}^\perp \left(\mathbf{r} - \mathbf{r}';s\right) [{s{\hat{\boldsymbol{{\mathcal{P}}}}}(\rb';s)-\hat{\mathbf{P}}^{(S)}(\mathbf{r}')}]
\end{equation}
where ${\hat{\boldsymbol{{\mathcal{P}}}}}(\rb;s)$ is the Laplace transform of the polarization density field operator. 

In Appendix B, we have expressed $\overleftrightarrow{G}^\perp $ in terms of the retarded scalar Green function for the vacuum. 
The substitution of \ref{eq:GreenT10} in \ref{eq:temp} gives
\begin{multline}
\label{eq:temp1}
    {\hat{\boldsymbol{C}}}(s; 
    \rb,\rb')=-\mu_0s {{{G}}} \left(\rb-\rb';s\right) [{s{\hat{\boldsymbol{{\mathcal{P}}}}}(\rb';s)-{\hat{\mathbf{{{P}}}}}^{(S)}(\rb')}]+\\
    \frac{1}{\varepsilon_0 s}\nabla_r\nabla_r[G(\rb-\rb';s)- G(\rb-\rb';s=0)][s{\hat{\boldsymbol{{\mathcal{P}}}}}(\rb';s)-{\hat{\mathbf{{{P}}}}}^{(S)}(\rb')].
\end{multline}
In the time domain, this expression becomes
\begin{multline}
\label{eq:temp2}
    {\hat{\mathbf{c}}}(t; \rb,\rb')=-\mu_0\frac{\partial}{\partial t}\int_0^t {{g}}\left(\mathbf{r} - \mathbf{r}';t-\tau\right) \dot{\hat{\mathbf{P}}}(\mathbf{r}';\tau)d\tau+\\
    \frac{1}{\varepsilon_0}\nabla_r\nabla_r\int_0^t [g(\rb-\rb';t-\tau)- g_0(\rb-\rb';t-\tau)]{\hat{\mathbf{{P}}}}(\rb';\tau)d\tau- \\
    \frac{1}{\varepsilon_0}\nabla_r\nabla_r \int_0^t [g(\rb-\rb';t-\tau)-g_0(\rb-\rb';t-\tau)]u(\tau){\hat{\mathbf{{{P}}}}}^{(S)}(\rb')d\tau,
\end{multline}
where
\begin{equation}
    {g}\rt=\frac{1}{4\pi r}\delta(t-r/c_0)
\end{equation}
and
\begin{equation}
    {g}_0(\rb;t)=\frac{1}{4\pi r} \delta(t).
\end{equation}
Using \ref{eq:temp2}, the solenoidality of the polarization density field operator in $V$, and the Gauss theorem, the expression \ref{eq:Etime10} is rewritten as
\begin{multline}
\label{eq:EApp}
  \hat{\mathbf{E}}\rt = -\mu_0\frac{\partial}{\partial t}\int_V\frac{{\dot{\hat{\mathbf{P}}}}(\mathbf{r}';t')}{4\pi\left| {\bf r} - {\bf r}' \right|}u(t')\dV' - \\ \frac{1}{\varepsilon_0}\nabla\int_{\partial V} \,\frac{{\hat{\mathbf{P}}}(\mathbf{r}';t')\cdot\,\mathbf{n}(\mathbf{r'})}{4\pi\left| {\bf r} - {\bf r}' \right|} u(t')\dS'+\hat{\mathbf{E}}_{f}
\end{multline}
where
$t' = t-|\mathbf{r}-\mathbf{r}'|/c_0$, $u(t)$ is the Heaviside function,
\begin{equation}
  \hat{\mathbf{E}}_{f}=\hat{\mathbf{E}}_{f}^\perp+\hat{\mathbf{E}}_{f}^\parallel
\end{equation}
and
\begin{equation}
    \hat{\mathbf{E}}_{f}^\parallel\rt = -\frac{1}{ \varepsilon_0}\nabla \int_{\partial V}  \frac{{\hat{\mathbf{P}}}^{(S)}(\mathbf{r}')\cdot \mathbf{n}'\rp}{4\pi |\mathbf{r}-\mathbf{r}'|}[u(t)-u(t')]\dS'.
\end{equation}
The term of ${\hat{\mathbf{c}}}(t; \rb,\rb')$ depending on $\hat{\mathbf{P}}(\mathbf{r}';t')$ through ${g}_0(\rb)$ cancels the longitudinal component of the electric field $\mathbf{E}^\parallel\{\hat{\mathbf{P}}\}$.

The \textit{free electric field operator} $\hat{\mathbf{E}}_{f}$ has two components, the transverse component $\hat{\mathbf{E}}_{f}^\perp$ that only depends on the free vector potential operator and the longitudinal component $\hat{\mathbf{E}}_{f}^\parallel$ that only depends on the initial value of the free polarization density field operator.

\subsection{Magnetic field operator}

The magnetic field operator is given by
\begin{equation}
\label{eq:Btime}
    \hat{\mathbf{B}}\rt =\mu_0 \nabla \times \int_V\dV' {\overleftrightarrow{g}}^\perp \left(\mathbf{r} - \mathbf{r}';t\right) * \dot{\hat{\mathbf{P}}}(\mathbf{r}';t')  + \hat{\mathbf{B}}_f
\end{equation}
where $\hat{\mathbf{B}}_f=\nabla \times \Af$.
Using \ref{eq:GreenT10} we obtain in the time domain
\begin{equation}
    \hat{\mathbf{B}}\rt =\mu_0 \nabla \times \int_V\frac{{\dot{\hat{\mathbf{P}}}}(\mathbf{r}';t')}{4\pi\left| {\bf r} - {\bf r}' \right|}u(t')\dV' + \hat{\mathbf{B}}_f\rt
\end{equation}
where $\hat{\mathbf{B}}_f=\nabla \times \Af$.

\section{Integral equation \ref{eq:IntegralEquation} in the frequency domain}
\label{sec:Frequency}
Equation \ref{eq:IntegralEquation} can be solved in the frequency domain, and then the solution in the time domain is evaluated using the inverse Fourier transform. Applying the Laplace transform to both sides of equation \ref{eq:IntegralEquation} we obtain the integral equation 
\begin{equation}
\label{eq:PolF1}
 \frac{1}{\Tilde{\chi}(s)} \Phat_\beta^{(\alpha)} (\rb;s) -\varepsilon_0\mathcal{L}_s\{ \Phat_\beta^{(\alpha)} \} = \Dhat_\beta^{(\alpha)} (\rb;s),
\end{equation}
where ${\hat{\boldsymbol{{\mathcal{P}}}}}_\beta^{(\alpha)}(\rb;s)$ is the Laplace transform of $\boldsymbol{p}_\beta^{(\alpha)}\rt$, $\Tilde{\chi}(s)$ is given by \ref{eq:chiTilde},
\begin{multline}
\label{eq:Funct}
  \mathcal{L}_s\{ \Phat_\beta^{(\alpha)} \}(\rb;s)=-\frac{s^2}{c_0^2}\int_V G(\rb-\rb';s) \Phat_\beta^{(\alpha)}  (\rb';s) \dV' - \\ \nabla\int_{\partial V} \,G(\rb-\rb';s) \Phat_\beta^{(\alpha)}  (\rb';s) \cdot\,\mathbf{n}(\mathbf{r'}) \dS'.
\end{multline}
The vector field $\Dhat_\beta^{(\alpha)} (\rb;s)$ for $\alpha = rad$, $\beta =\mu$ is given by
\begin{equation}
\Dhat_\mu^{(rad)} (\rb;s)= \varepsilon_0 \mathcal{E}_\mu \left(\frac{1}{s+i\omega_\mu}+\frac{i}{\omega_\mu} \right) \mathbf{w}_\mu\rp,
\end{equation}
and for $\alpha = mat$, $\beta =m$ is given by
\begin{equation}
\boldsymbol{\mathcal{D}}_{m,\nu}^{(mat)}\rs=
\sqrt{\frac{\hbar \sigma(\nu)}{\nu \pi}}\left[\frac{1}{{\Tilde{\chi}}(s)} \frac{\mathbf{U}_m(\mathbf{r})}{s+i\nu} +\boldsymbol{\mathcal{M}}_m(\mathbf{r};s)\right],
\end{equation}
where
\begin{equation}
\boldsymbol{\mathcal{M}}_m \rs =\frac{s}{c_0^2}\int_V {G}(\rb-\rb';s) \mathbf{U}_m(\mathbf{r}')\dV'+\boldsymbol{\mathcal{N}}_m \rs 
\end{equation}
and
\begin{equation}
\boldsymbol{\mathcal{N}}_m \rs =-
\frac{1}{\varepsilon_0 s}\nabla \int_{\partial V} [{G}(\rb-\rb';0) -{G}(\rb-\rb';s)]\hat{\mathbf{U}}_m(\mathbf{r}')\dS'.
\end{equation}
The first term in the expression of $\boldsymbol{\mathcal{M}}_m$ is due to the initial condition $\boldsymbol{p}^{(mat)}_{m,\nu}(\mathbf{r};t=0)=\mathbf{U}_m(\mathbf{r})\sqrt{\frac{\hbar \sigma(\nu)}{\nu \pi}}$. 

It is sufficient to solve equation \ref{eq:PolF1} for $s=i\omega+0^+$ with $-\infty<\omega<\infty$, so it reduces to a volume integral equation in the frequency domain. To deal with the singularities at $\omega=\pm\omega_\mu$ and $\omega=\pm\nu$ we use the relation \ref{eq:Princv}.

\end{document}